\documentclass[a4paper]{article}

\usepackage[parfill]{parskip}
\usepackage[pdf]{graphviz}
\usepackage[utf8]{inputenc}
\usepackage[toc,page]{appendix}
\usepackage{glossaries}
\usepackage{algorithm}
\usepackage{algpseudocode}
\usepackage{amssymb}
\usepackage{graphicx}
\usepackage{subcaption}
\usepackage{booktabs}
\usepackage{amsmath}
\usepackage{tikz}
\usepackage{standalone}
\usepackage[disable]{todonotes}
\usepackage{geometry}

\usepackage{algorithmicx}
\usepackage{algpseudocode}
\algdef{SE}[DOWHILE]{Do}{doWhile}{\algorithmicdo}[1]{\algorithmicwhile\ #1}

\usepackage{booktabs}
\usepackage{tabularx}

\PassOptionsToPackage{hyphens}{url}
\usepackage{hyperref}
\usepackage{cleveref}

\DeclareMathOperator*{\argmax}{arg\,max}

\usetikzlibrary{shapes,arrows}

\tikzstyle{block} = [rectangle, draw, 
    text width=3em, text centered, minimum height=3em]
\tikzstyle{round} = [circle, draw, 
    text width=3em, text centered, minimum height=3em]

\title{A Unifying Hybrid Consensus Protocol}

\author{
  Yulong Wu\\
  \texttt{yulong@aion.network}
  \and
  Yunfei Zha\\
  \texttt{frederic@aion.network}
  \and
  Yao Sun\\
  \texttt{yao@aion.network}
}

\date{v0.1.1 - \today}

\begin{document}

\maketitle

\begin{abstract}
We introduce Unity, a new consensus algorithm for public blockchain settings. Unity is an eventual consistency protocol merging the Proof-of-Work (PoW) and Proof-of-Stake (PoS) into a coherent stochastic process. It encompasses hardware and economic security without sacrificing availability, unpredictability and decentralization. Empirical results indicate that the proposed protocol is fair and scalable to an arbitrary number of miners and stakers.
\end{abstract}
\clearpage

\tableofcontents
\clearpage

\section{Introduction}

Eventual Consistency based consensus protocols have shown incredible resilience over the past decade as it has until recent years remained a persistent standard in many popular network implementations. As explored by Miller and La Viola \cite{miller2014anonymous}, PoW consensus (a.k.a. Nakamoto consensus) protocols adhere well in public settings. These protocols expose no message complexity to limit the size of the network, nor do they expose any interactive membership system to constrain the validator set. Instead, PoW offers a non-interactive puzzle based purely on the computational power of all participants in the network. Therefore the security of the network depends \textit{mostly} on the relative computational power of the malicious party.

While these protocols do provide consensus, they don't necessarily provide immutability as by their very nature they are designed to reach consensus that favours the total work done by parties in the network. One flaw in the system as a whole is the expected distribution of computation resources within the network. Whereas it was initially assumed to be a flat, mostly uniform distribution, the rise of mining pools \cite{aliaga2018} and ASICs \cite{monerocrusher2019} leave many initial assumptions of resource breakdown mostly obsolete. In recent years it has become apparent that the disparity in computational resources between larger and smaller networks and the increasing efficiency of resource renting services revokes previous assumptions of the impracticality of a double-spend attack in smaller networks \cite{sinnige2018}.

An alternative approach known as Proof-of-Stake (PoS) encompasses a broad spectrum of consensus protocols that share the common attribute in utilizing the \textit{stake} a user has in a network \cite{BentovGM14}. Proponents here argue that utilizing stake has the inherent benefit of better aligning the incentives of the user and network, as well as providing a protocol designer more flexibility in punishing behaviour not intended by the protocol (slashing) \cite{ethereum1}. Pure PoS also suffers from criticisms. Specifically, it changes the security model to be based economic activities, introducing the possibility of coin swing attacks and increasing the feasibility for plutocracy and collusion \cite{buterin_plutocracy}.

In this paper, we propose a novel consensus protocol, Unity, which merges pure PoW and PoS into a coherent stochastic process, which demonstrates the highest level of security and strongest resilience to economic exploits.

\subsection{Contributions}

This document will present:

\begin{itemize}
    \item A novel hybrid consensus protocol called Unity that retains the non-interactivity originally obtained from Nakamoto consensus and places little restrictions on the network size.
    \item Results from simulations at varying degrees of details, providing an argument for the long term sustainability and coherence of such a protocol.
\end{itemize}

\section{Related Works}

Proof-of-Stake (PoS) algorithms first appeared as little more than a thought experiment on the Bitcoin Forum \cite{bitcointalk2011pos}. The work of Popov \cite{popov2016probabilistic} demonstrated the ability to formulate a pseudo-random deterministic ordered list, and the demonstration that such a mechanism could be used to effectively simulate a stochastic process similar to that of PoW. Of great interest to this paper is the fact that the exact order of the list is not known unless all players choose to publish their respective ranks. Working off several previous designs, Jepson \cite{jepson2015} demonstrated the ability to mitigate private mining by forcing increasingly expensive PoW mining via modifications to the fork choice rule to account for the stake within the network.

\section{Preliminaries}

Assume the existence of a network of interconnected nodes. Each node maintains a state $\delta$ that contains a tree of homogeneous blocks, and that blocks are linked through hashes. Define $\mathbb{N}_{x}$ the set of all non-negative integers smaller than $2^x$. Define a block $b = (f_{p}, f_{sr}, f_{tr}, f_{d}, f_{ts}, f_{tx})$, where $f_{p}, f_{sr}, f_{tr}, f_{d} \in \mathbb{N}_{256}$, $f_{ts} \in \mathbb{N}_{64}$, $f_{tx}$ is a list. The exact purpose of each element is defined in \cref{tab:block}.

\begin{table}[h]
    \centering
    \begin{tabularx}{0.8\textwidth}{l X}
        \toprule
        \textbf{\emph{Element}} & \textbf{\emph{Description}} \\
        \midrule
        $f_{p}$ & The hash of the parent block.\\
        $f_{sr}$ & The hash of the root node of the state trie, after all transactions are executed and finalisations applied.\\
        $f_{tr}$ & The hash of the root node of the trie structure populated with each transaction in the transactions list portion of the block.\\
        $f_{d}$ & The mining and/or staking difficulty.\\
        $f_{tx}$ & The included transactions.\\
        $f_{ts}$ & The timestamp of the block generation.\\
        \bottomrule
    \end{tabularx}
    \caption{The basic elements of a block}
    \label{tab:block}
\end{table}

There exists a set of users that are participating on the network. Refer to this as the universe of the blockchain and denote the set of all users by $\xi = \{x \mid x \in \mathbb{N}^{0} \wedge x \leq 2^{256}\}$, where $x$ represents a unique ID corresponding to the account. Assume then there are two other sets $H$ and $B$ that refer to the active computational power, and balance of the user. There exists $f_{hp} : \xi \longmapsto H$, $f_{balance} : \xi \longmapsto B$, now define the set $M = \{\mu \in \xi \mid f_{hp}(\mu) \neq 0\}$, as the set of active miners, and define $S = \{s \in \xi \mid f_{balance}(s) \neq 0 \}$, these form the two main parties of interest in the system. $S$ represents all actively participating stakers (implicit, that all users with balance are participating, this can be mapped onto scenarios where only a proportion of the network is participating in staking, see \cite{popov2016probabilistic}).

A transaction represents a valid arc between two states, formally:
$$
    \delta_{t+1} \equiv \Upsilon(\delta_t, T),
$$
where $\Upsilon$ is the state transition function and $T$ is a transaction. $\Upsilon$ can be as simple as a balance transfer function or more powerful to include arbitrary state transactions through a virtual machine.

\section{Unity - A Hybrid PoW/PoS Consensus}

The design of Unity is relatively (and deliberately) simple. The design is such that there are two relatively independent stochastic processes. Here the novelty is mainly in that the security of the system relies both on computational resources and active stake of the network. The game being played is similar to that of PoW: all active miners (who produce PoW blocks) and stakers (who produce PoS) blocks compete in two separate lotteries respectively. The first to find a new block publishes and obtains rewards for doing so. We define a fork selection rule based on an intermediate unit that incorporates both total computational resources and total active stake.

Also similar to PoW is a network parameter, defined beforehand, and two separate difficulty parameters (for PoW and PoS) that fluctuate (see \cref{diff_adj}), to maintain the parameter.

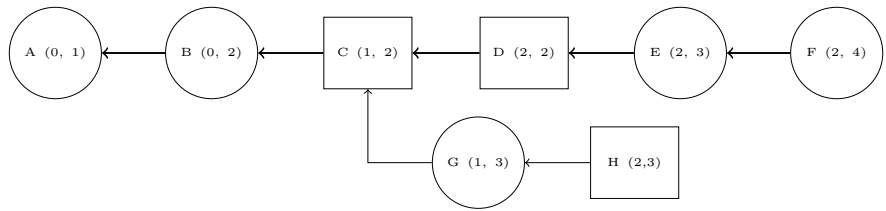
\begin{figure}[h]
    \centering
    \begin{tikzpicture}[node distance = 6.5em, auto]
    \node (b1) [round] {\tiny A (0, 1)};
    \node (b2) [round, right of=b1] {\tiny B (0, 2)};
    \node (b3) [block, right of=b2] {\tiny C (1, 2)};
    \node (b4) [block, right of=b3] {\tiny D (2, 2)};
    \node (b5) [round, right of=b4] {\tiny E (2, 3)};
    \node (b6) [round, right of=b5] {\tiny F (2, 4)};
    
    \node (c1) [round, below right of= b3] {\tiny G (1, 3)};
    \node (c2) [block, right of= c1] {\tiny H (2,3)};

    \draw [<-, thick] (b1) -- (b2);
    \draw [<-, thick] (b2) -- (b3);
    \draw [<-, thick] (b3) -- (b4);
    \draw [<-, thick] (b4) -- (b5);
    \draw [<-, thick] (b5) -- (b6);
    
    \draw [<-] (b3) |- (c1);
    \draw [<-] (c1) -- (c2);
    
\end{tikzpicture}
    \caption{A depiction of a Unity chain, note that there is no enforced pattern or ordering on PoW blocks (Square) and PoS blocks (Circle). Note that $(td_w, td_s)$ represents the total mining and staking difficulty respectively. In this scenario, the bold arrows represent the canonical chain. As explained later in \cref{sec:fork_choice_rule}, this is due to have a higher combined total difficulty of $td=td_w \cdot td_s=2 \cdot 4=8$ vs $6$.}
    \label{fig:high_level_unity}
\end{figure}

The protocols assumes that honest nodes in the network will always mine on the heaviest known chain. In the standard case the first to publish a new block has the highest chance of being included into the main chain, as it is most likely to be seen by the majority of nodes in the network. It is also obvious to see that the two stochastic processes must appear to be i.i.d, otherwise we could not assume the behaviour of honest nodes to be optimal (for example imagine if the PoS process was not memoryless, this would drastically change the behaviour of the game). A simple depiction of a main chain and a fork is shown in \cref{fig:high_level_unity}.

The next few sections provides details on the block forging processes for participants, and makes an argument via simulation for why they can be treated as i.i.d.

\subsection{Block Forging}

In this setup, miners working on a PoW will operate in a similar fashion to a conventional mining setup, where a naive (but concise) implementation would have the miner performing the following steps as shown in \cref{mining_algo}; one can see that the game being played is still the same. Assume the mining difficulty parameter $d_w$ and the existence of a 256-bit hash function $hash(\cdot)$, the miners solve the following cryptographic puzzle:
\begin{equation}
hit = hash(b) \leq 2^{256} / d_w.
\end{equation}
Treat $hash(\cdot)$ as a random oracle with uniform distribution in $\mathbb{N}_{256}$. There is some reward that is distributed to the miners for successfully mining a block. As the cryptographic puzzle is independent of the miner, the probability of a miner mining a block is proportional to the number of permutations a user can to evaluate, its computational power.

\begin{algorithm}
    \caption{Correct Mining Loop, with memory pool (pending state), and a difficulty adjustment function.}
    \label{mining_algo}
    \begin{algorithmic}[1]
        \Procedure{MINEBLOCK}{$\delta$}\Comment{The mining loop}
        \State $c    \gets \Call{GetBestChain}{}$
        \State $b_1  \gets \Call{GetLastPowBlock}{c}$
        \State $b_2  \gets \Call{GetSecondLastPowBlock}{c}$
        \State $diff \gets \Call{GetDifficulty}{b_1, b_2}$
        \State $txs  \gets \Call{GetMemPoolTxs}$
        \State $b    \gets \Call{CreatePowBlockTemplate}{c, txs}$
        \Do
            \State $solution \gets \Call{PoW}{b}$
        \doWhile{$solution > 2^{256}/diff$}
        \State $b    \gets \Call{Finalize}{b, solution}$
        \State $\Call{ImportAndPropagate}{b}$
        \EndProcedure
  \end{algorithmic}
\end{algorithm}

This is in contrast to block forging in a PoS setup, in which we derive a mostly non-interactive forging protocol first formally proposed by Popov \cite{popov2016probabilistic}, and later expanded upon by Begichava and Kofman \cite{begichava2018} to provide fair distribution. The proposed PoS forging algorithm here follows close to \cite{begichava2018}, and simulations show the distribution to be fair relative to staking power. More importantly, this implementation is \textit{not} the transparent forging scheme described in later versions of \cite{popov2016probabilistic}. Thus, it preserves the nice characteristic of not revealing the rank ordering of the list for any party unless the party chooses to submit a block.

In the PoS forging described in \cref{staking_algo}, every staker computes a deterministic pseudo-random number, $seed$, using its public-private key pair $(pk, sk)$,
\begin{equation}
seed_{t + 1} = sign(seed_t, sk).
\end{equation}
For the given staker, a PoS block can be produced if and only if
\begin{equation}
|\ln({hash(seed)}/{2^{256}})| \cdot d_s \leq V \cdot \Delta,
\end{equation}
where $V$ is the amount of stake (or voting power) and $\Delta$ is the time elapsed since the last PoS block. The $\Delta$ can also be computed by solving
\begin{equation}
\Delta \geq \frac{d_s \cdot \ln(hash(seed)/2^{256})}{V}.
\end{equation}

\begin{algorithm}
    \caption{Correct Staking Forge Loop, here $V$ refers to the voting power of the account, $\phi$ refers to the current timestamp of the forger.}
    \label{staking_algo}
    \begin{algorithmic}[1]
        \Procedure{FORGEBLOCK}{$\delta, pk, sk$}
        \State $c     \gets \Call{GetBestChain}{}$
        \State $b_1   \gets \Call{GetLastPosBlock}{c}$
        \State $b_2   \gets \Call{GetSecondLastPosBlock}{c}$
        \State $stake \gets \Call{GetStake}{c, pk}$
        \State $diff  \gets \Call{GetDifficulty}{b_1, b_2}$
        \State $ts    \gets \Call{GetTimestamp}{b_1}$
        \State $seed  \gets \Call{GetSeed}{b_1}$
        \State $seed  \gets \Call{Sign}{seed, sk}$
        \State $\Delta \gets diff \cdot \ln(hash(seed)/2^{256}) / stake$
        \Do
            \State $\Call{Sleep}{1}$
        \doWhile{$\phi < ts + \Delta$}
        \State $txs \gets \Call{GetMemPoolTxs}$
        \State $b \gets \Call{CreatePosBlockTemplate}{c, txs, seed}$
        \State $b \gets \Call{Finalize}{b, sk}$
        \State $\Call{ImportAndPropagate}{b}$
        \EndProcedure
    \end{algorithmic}
\end{algorithm}

\subsubsection{Voting Power and Active Stake}

\begin{figure}[h]
    \centering
    \includegraphics[scale=0.5]{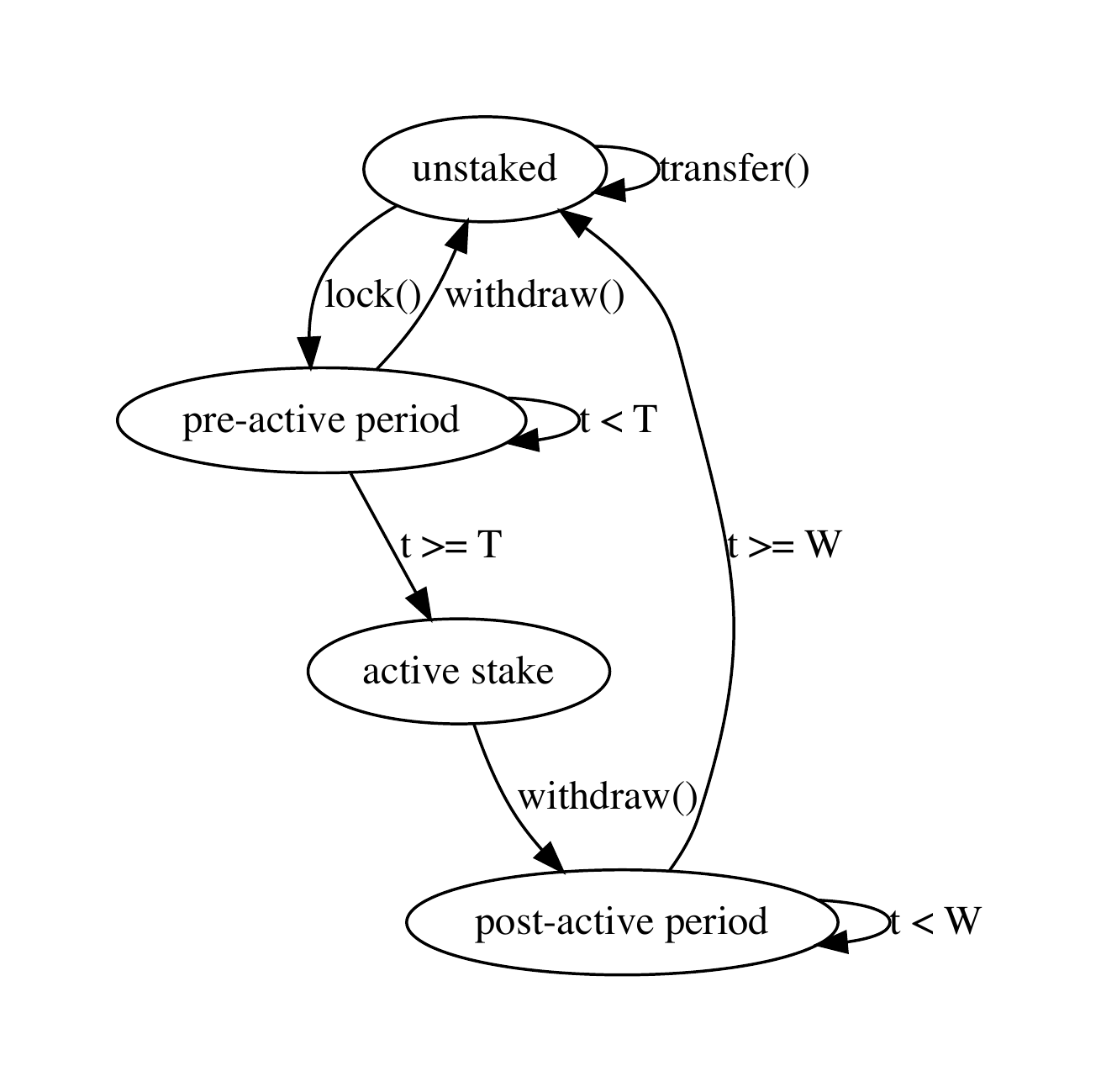}
    \caption{Possible states for coins in an account. Where $T$ refers to a constant specifying the time (in height) it takes for stake to become active. $W$ refers to a constant specifying the time it takes for stake to become unlocked.}
    \label{fig:stake_state}
\end{figure}

Voting power $V$ is the active stake for a particular account. The balance of every account is initially unstaked, and therefore does not contribute to $V$. The user must called the $lock()$ function to lock balance. There is then a cycle of maturation that occurs before locked stake becomes active. Note that the concept of maturity is boolean, the locked coins are either matured (in which case they are active stake) or not (see \cref{fig:stake_state}). This is to prevent any staking grinding attacks (see \cref{sec:stake_grinding}) and balance shifting attacks \cite{nxt_wiki}. The withdrawal period lock-up is to facilitate a possible slashing mechanism (see Appendix \ref{appendix:slashing}).

\subsection{Difficulty Adjustment}
\label{diff_adj}

As discussed in \cite{bitfury2015}, both PoW and PoS block generation time is distributed exponentially. In case of PoW, the rate $\lambda_w$ is equal to $r/d_w$, where $r$ is the hash rate and $d_w$ is the mining difficulty; in PoS, the rate $\lambda_s$ is equal to $b/d_s$, where $b$ is the stake and $d_s$ is the staking difficulty.

The goal of difficulty adjustment algorithm is to achieve a mean target block time of $t$. For a random variable $X \sim Exp(\lambda)$, the mean is $\lambda^{-1}$. Let $\lambda_w^{-1} = 2t$ and $\lambda_s^{-1} = 2t$, then $d_w = r \cdot 2t$ and $d_s = b \cdot 2t$. However, the network hash rate can not be determined by the protocol and changes frequently as miners come and go; the network stake can be calculated but the active stake is unknown due to some of the nodes being offline. Therefore, the difficulties can only be learned.

Ideally, a difficulty adjustment algorithm should have the following common properties:
\begin{itemize}
    \item \textbf{Responsiveness} -- The algorithm should respond quickly to changes in the network hash rate or stake power.
    \item \textbf{Low volatility} -- The difficulty should be relatively stable when the network hash rate or stake power is constant. 
    \item \textbf{Simplicity} -- The algorithm should be relatively simple to implement.
    \item \textbf{Low memory} -- The algorithm should not require a memory of many past block timestamps.
\end{itemize}

In Unity, a new algorithm is proposed. Let $2t$ be the target time (between PoW blocks and between PoS blocks), the algorithm works in the following way:
\begin{itemize}
    \item $d := 1$;
    \item let $X$ be an exponential random variable with rate $\lambda = \frac{1}{2t}$;
    \item if $X > -\frac{\ln (0.5)}{\lambda}$, then $d_{t+1} = \frac{d_t}{1 + \alpha}$;
    \item if $X = -\frac{\ln (0.5)}{\lambda}$, then $d_{t+1} = d_t$;
    \item if $X < -\frac{\ln (0.5)}{\lambda}$, then $d_{t+1} = d_t \cdot (1 + \alpha)$.
\end{itemize}
The $\alpha$ controls the learning rate, which further determines the responsiveness of the algorithm. The $-\frac{\ln (0.5)}{\lambda}$ is calculated by solving the following equation,
\begin{equation}
    1-e^{(-\lambda x)}=0.5,
\end{equation}
where the left-hand side is the cumulative distribution function of an exponential random variable.

Note that the difficulty adjustment algorithm is primarily based on block timestamp. The honest nodes are required to not process a future block. See more detailed analysis in \cref{future_mining_attack}.

\subsection{Fork Choice Rule}
\label{sec:fork_choice_rule}

The Unity fork choice rule is based on the product of total mining difficulty $td_w$ and total staking difficulty $td_s$,
\begin{equation}
c_0 = \argmax_{i \in \{1,...,N\}}{td_{wi} \cdot td_{si}},
\end{equation}
where $c_0$ is the canonical chain and $N$ is total number of forks.

\subsection{Preservation of Nakamoto Consensus Traits}

Many of the common proofs of effectiveness of Nakamoto Consensus have relied on the treatment of the stochastic process as a Poisson process. Nakamoto's initial introduction of Bitcoin \cite{nakamoto2008} assumes this when calculating the feasiility of an attack, and Eyal and Sirer make the same assumption in derivations of the selfish mining attack \cite{eyal2018majority}. Both the traditional PoW and PoS mechanisms described in Unity adhere to these principles, therefore assuming that both mechanisms are i.i.d (which the difficulty function should provide) we can argue that:

\begin{equation}
    P(Z) = \frac{(\lambda_x + \lambda_y)^z}{z!}e^{-(\lambda_x+\lambda_y)}
\end{equation}

Follow the proof in \cite{pishro2016introduction}. Let $N_x(t)$ and $N_y(t)$ be two independent Poisson processes with rates $\lambda_x$ and $\lambda_y$ respectively. Let us define $N(t)=N_x(t)+N_y(t)$. That is, the random process $N(t)$ is obtained by combining the arrivals in $N_x(t)$ and $N_y(t)$. First, note that
\begin{equation}
    N(0)=N_x(0)+N_y(0)=0+0=0.
\end{equation}
Now, consider an interval of length $\pi$, i.e, $I=(t,t+\pi]$. The numbers of arrivals in $I$ associated with $N_x(t)$ and $N_y(t)$ are $Poisson(\lambda_x \pi)$ and $Poisson(\lambda_y \pi)$ and they are independent. Therefore, the number of arrivals in $I$ associated with $N(t)$ is $Poisson((\lambda_x+\lambda_y) \pi$).

A well known property which states that the summation of two Poisson RVs is another Poisson RV in which the rate is the summation of the two. The simulations show that this result is indeed correct, as $\lambda_{PoS,PoW}=\lambda/2$. Therefore the target rate as we expected was $\lambda_{PoS,PoW} \cdot 2 = \lambda$.

\subsection{Slashing}

Whether slashing should be enforced is left to the implementation. For an analysis over different slashing protocols, see Appendix \ref{appendix:slashing}.

\section{Security Analysis}

We examine attacks published in past literature \cite{buterin_randomness}\cite{ouroboros}.

\subsection{Private Double-spend Attack}

Conventional Nakamoto consensus is subject to double-spend attacks, because transactions are irreversible and forks are allowed. For instance, in Bitcoin, if an attacker controls 51\% of the total hash power, he can produce a side chain with a higher total difficulty than the current main chain. After a spending transaction has been acknowledge by a merchant or exchange, the attacker can release a side chain to reverse that transaction. Similarly, an attacker can launch a double-spend attack if he controls over 51\% of the total stake, for a pure PoS network.

In Unity, it is much more difficult for an attacker with 51\% hash power or 51\% stake power to launch a similar attack, because the fork choice rule takes both mining and staking into account. It would require a combination of hash and stake power of over 100\% to dominate the system.

\todo{Needs formal analysis}The fork choice rule also implies that one's power is maximized only when he possesses close mining power and staking power. Unity's hybrid consensus balances the power between mining giants and big coin holders in such a way that not a single part can harm the security of the system. Additionally one could argue that the incentives for launching this attack as a stake, and thus coin holder becomes less plausible as the net drop in price due to fallout impacts all coin holders negatively.

\subsubsection{Calculations}

Consider the scenario where an attacker tries to generate an alternative chain in private, and reveal it at a later point. Assume the attacker has a hash power and stake power $(a, b)$; and the honest nodes has $(c, d)$.

Based on the analysis in \cite{bitfury2015}, the PoW block generation rate $\lambda_w = \frac{w}{d_w}$, where $w$ is the hash rate. In a period of unit time, the number of blocks will be produced is a random variable $X \sim Pois(\lambda_w)$, and $E(X) = \lambda_w$. Let $Y_w$ be the total mining difficulty, then $E(Y_w) = E(X) * d_w$. Thus,
$$
E(Y_w) = \frac{w}{d_w} * d_w = w.
$$

Similarly, the PoS block generation rate $\lambda_w = \frac{s}{d_s}$, where $s$ is the stake. Let $Y_s$ be the total mining difficulty, then
$$
E(Y_s) = \frac{w}{d_s} * d_s = s.
$$

Thus, the total mining difficulty is an integration of hash rate over time, while the total staking difficulty is an integration of stake over time. Given a time duration $t$, the attacker's chain has an expected weight of
$$
(td_{wc} + a \cdot t) \cdot (td_{sc} + b \cdot t)
$$
and the honest nodes' chain has
$$
(td_{wc} + c \cdot t) \cdot (td_{sc} + d \cdot t),
$$
where $td_w$ and $td_s$ are the total mining difficulty and total staking difficulty from the genesis block to the fork point.

For the attacker to overthrow the honest nodes' chain, the attacker's chain has to own a higher weight than the honest nodes' chain, which further leads to the following inequality:
\begin{equation}
\label{eq}
    td_{sc} \cdot (a - c) + td_{wc} \cdot (b - d) + (ab - cd) \cdot t \geq 0.
\end{equation}

\subsection{Nothing-At-Stake}
\label{sec:nothing_at_stake}

Nothing-At-Stake is a well known issue of PoS blockchains where staking has literally no cost. Coin holders can stake on every branch they see to maximize their profits no matter which branch wins the fork competition. This harms the system in a way that forks may be maintained for a long time (the finality is delayed) and, attackers only need 1\% stake power to select a winner fork to conduct double-spending.

Some solutions try to punish accounts who staked on multiple branches but a strategy called undetectable nothing-at-stake\cite{brown2018formal} allows attackers to enact this without leaving hard evidence. This attack amounts to splitting stake into multiple accounts, the net probability of the set being validated remaining the same. No punishment can be enacted on the attacker in this scheme as it is impossible to prove (from the information observed by the network) two accounts are owned by the same individual.

Unity doesn't suffer from this problem thanks to its hybrid nature: Forks can be efficiently eliminated by PoW mining as miners will eventually concentrate on one fork just like in pure PoW systems.

\subsection{Long Range Attack}
\label{sec:lra}

Long Range Attacks (LRAs) refer to a scenario where the attacker reverts to a depth significantly distant (in the past) from the current time, in which the individual has compromised $>50\%$ of staking power in the network. This attack is only feasible when the creation of blocks is free, therefore assume that an attacker will try to launch this attack with only PoS blocks. Then, one can argue that,

\begin{equation}
\begin{split}
\begin{aligned}
td_{a} &= \sum_{i = 1..H_w - n}{d_{wi}} \cdot \sum_{j = 1...H_s}{d_{sj}} \\
&= ((H_w - n) \cdot \overline{d_{w}}) \cdot (H_s \cdot \overline{d_{s}}) \\
\end{aligned}
\end{split}
\end{equation}

Where $td_{a}$ represents the total difficulty of the attackers chain. It is trivial to see that even if the attacker were to own all of the active stake, assuming the total voting power does not change, the best case is an equal $td$ to the main chain. Because the protocol disallows future blocks, the expected upper bound blocks being forged by the LRA is $(\phi - t_{N_w - n}) / 2t$. If the attacker is able to grow his staking power through block rewards, the chance of success if maximized the earlier the point of attack. The attacker would require 

\begin{equation}
\label{eq:additional_power}
\begin{split}
\begin{aligned}
(H_s \cdot (\overline{d_{s}} + \Omega)) &> (H_w \cdot \overline{d_{w}}) \cdot (H_s \cdot \overline{d_{s}}) \\
\Omega &> \frac{H_w \cdot \overline{d_{w}}}{H_s}
\end{aligned}
\end{split}
\end{equation}

Assuming that the main chain has static forging power (difficulty does not change), we expect $N_w = N_s$, we define a parameter $\Omega$ to represent the additional amount of power (represented in difficulty) an attacker would require to match the main chain. From \cref{eq:additional_power}, $\Omega$ must be greater than the difficulty of the PoW chain. Further exploration is needed on the time it takes for an attacker to achieve the additional difficulty, the hypothesis is that this process of acquiring power gradually via block rewards occurs over a long period of time.

\subsubsection{Past Majority Attack}

In one scenario, the attackers are validators who in the past held majority. In a pure PoS setting these types of attacks require the existence of assisted synchronization to ensure that a user who is excessively desynchronized from the network obtains correct information. However \cref{sec:lra} and this section indicate that Unity works even in the absence of this party, as the hybrid approach ensures the creation of a pure PoS fork is impractical.

\subsection{Stake Grinding}
\label{sec:stake_grinding}

Stake grinding is a form of attack where the attacker obtains some control over the generation of a seed and is able to \textit{grind} solutions (in a fashion similar to PoW) until the seed determines a series of random numbers favourable for the attacker as examined by Buterin\cite{buterin_randomness}. An obvious example of this is obtaining a seed that allows the attacker to generate the majority of blocks for the next $n$ blocks in a sequence.

The attacker must have a variable to grind. In this design, the block hash is not a viable source for this, as the previous seed is directly related to the unique signature of the last staker. One possible avenue of grinding could be to distribute PoS to a large amount of accounts where the signatures of each account (assuming this occurs in a sequence) leads to a high probability of another being selected. To combat the protocol specifically requires a maturing period. Therefore any information the attacker has in the present (or near present) has no effect assuming that this delay is set properly.

\subsection{Denial of Service}

Denial of Service attacks occur when an attacker is aware of who will be engaging in block production \textit{a priori} to the event. This allows the attacker to identify the physical location of the block producer and remove the node via DOS. Unity doesn't suffer (assuming a fairly distributed voting set) to this problem as validators are randomly selected in both PoW and PoS, and no information about block production eligibility is leaked \textit{until} the blocks are published.

\subsection{Selfish Mining}

Selfish mining refers to a scenario where attackers try to mine more blocks than their hash power deserves to with a strategy that: If the attackers mined block $B_1$, they don't publish it so that they can start mining the next block $B_2$ earlier than other miners in the network. But they also take risks as other miner may create and publish another $B_1'$ at any time and run ahead of attackers' hidden $B_1$.

Selfish mining is more powerful if one can predict when others can mine a block so that the risk of holding a mined block is reduced. Unity doesn't suffer from this problem as block creation is random in both PoW and PoS, and no information about it is leaked until the blocks are published.

Unity's hybrid consensus makes selfish mining impractical as attackers may miss a PoS block when holding a mined PoW block.

\subsection{Transaction Denial}

Transaction denial attacks are a censorship attack where the attacker wants to prevent a transaction from being confirmed. By \cref{sec:fork_choice_rule}, the leader is randomly selected via a lottery, and therefore randomly selects a block proposer. Therefore as long as the transaction is propagated to the majority of mining nodes it cannot be denied.

\subsection{Eclipse Attack}

Eclipse attacks define when a nodes view of the network is compromised via receiving only incorrect information from it is peers. In our setup a node in this position would be assumed to be an attacker.

\section{Results}

In this section, empirical evaluation results from simulation are presented. For simplicity, we simulate the hash output as a uniformly distributed variable $X \sim U(0, 1)$, for both miners and stakers. For the difficulty adjustment algorithm, we set $\alpha = 0.01$. Additionally, we have $S = [160, 80, 40, 30, 20, 10, 10, 10, 10, 10]$ and $M = [16, 8, 4, 3, 2, 1, 1, 1, 1, 1]$; the numbers are picked to demonstrate how different measurements respond to both linear and exponential increases in hash/stake power. The target block time $t$ is 10 seconds, and a period of 30 days is simulated. For anyone who is interested, the full simulation code is available at \url{https://github.com/aionnetwork/unity/blob/master/simulations/unity_poisson_sim/Unity.ipynb}.

\subsection{Block Rewards Fairness}

First, the block rewards fairness is studied. In the Unity context, fairness means that the number of PoS blocks and the number of PoW blocks should be equal or very close. Additionally, for the miners, the block rewards should be proportional to their hash power (normalized by the total network hash power); for the stakers, the block rewards should be proportional to their stake power (normalized by the total network stake).

\begin{figure}
    \centering
    \begin{subfigure}[b]{0.45\textwidth}
        \includegraphics[width=\textwidth]{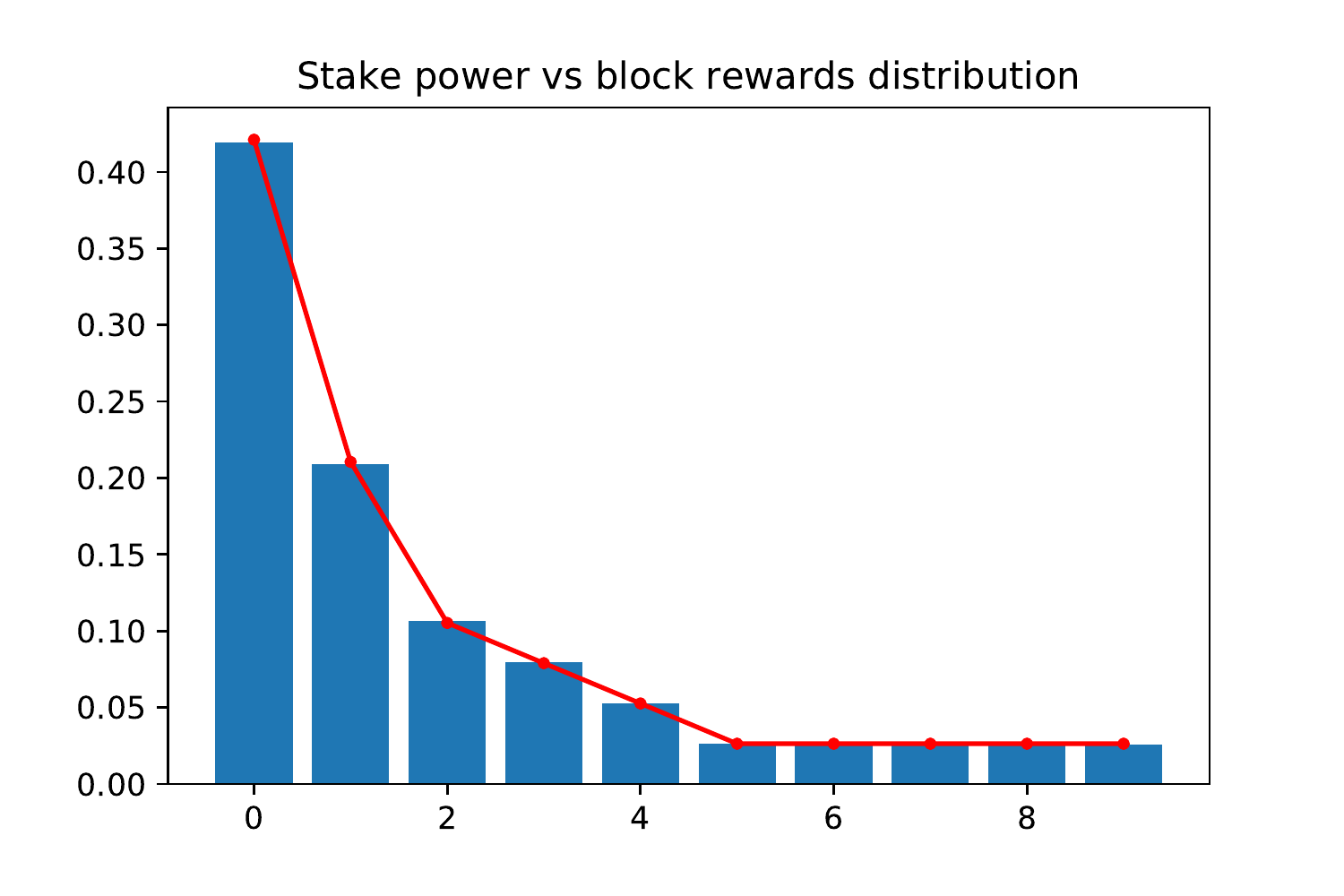}
        \caption{The stake power of each staker (bar) and the corresponding received block rewards (line).}
    \end{subfigure}
    \begin{subfigure}[b]{0.45\textwidth}
        \includegraphics[width=\textwidth]{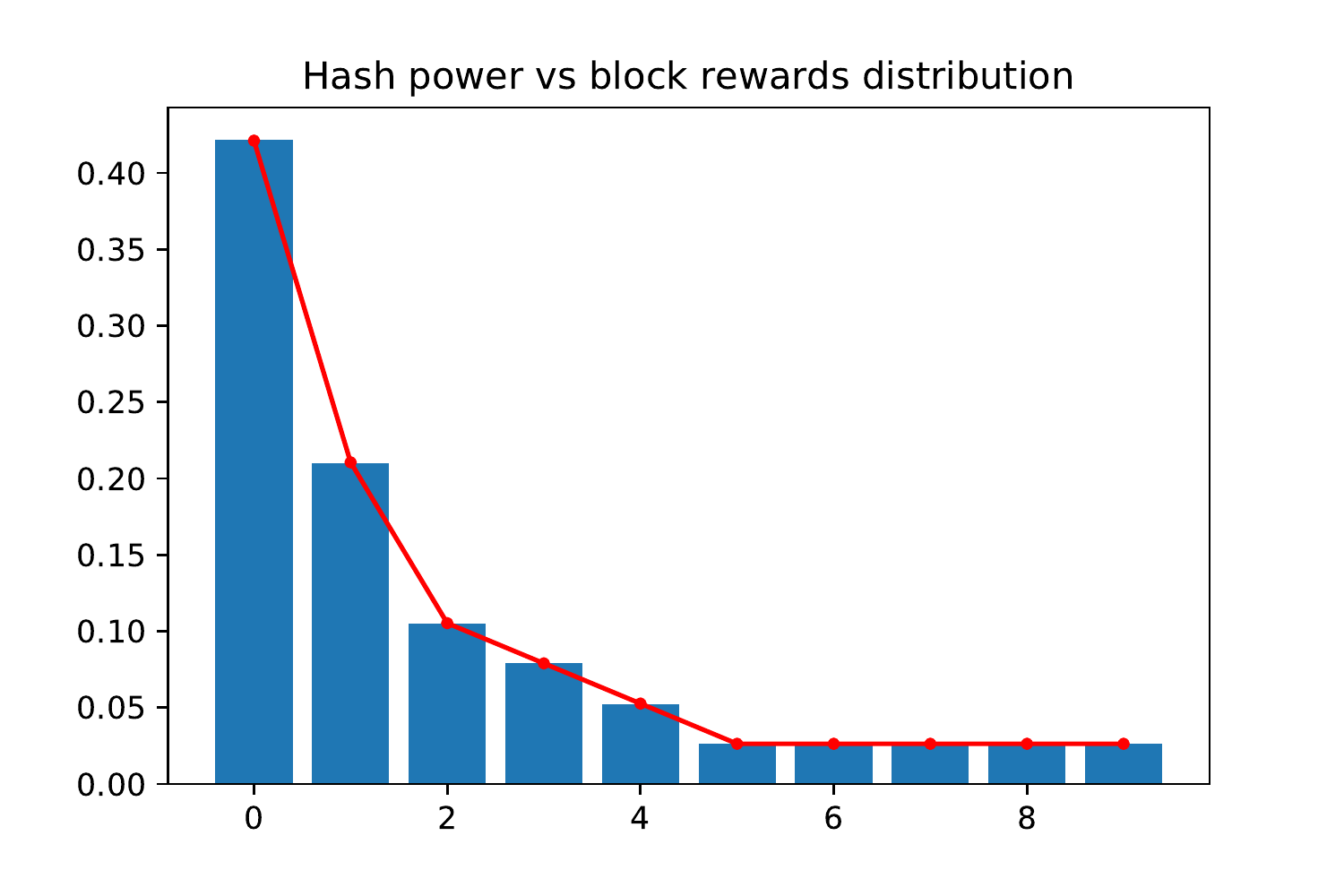}
        \caption{The hash power of each miner (bar) and the corresponding received block rewards (line).}
    \end{subfigure}
    \caption{Block rewards distribution vs relative hash/stake power.}
    \label{fig:block-rewards}
\end{figure}

During the simulated period, a total of 250044 blocks are generated, where 124858 of them are PoS blocks and 125186 of them are PoW blocks. The block reward that each miner/staker received (normalized) is depicted in \cref{fig:block-rewards}. Block rewards are observed to be proportional to the relative stake/hash power, thus the system can be deemed fair.

\subsection{Block Time Distribution}

Then, we look at the block time distribution based on our simulation. As described earlier, we want a target block time of 10 seconds, as a result of 20 second PoS block time and 20 second PoW block time.

\begin{figure}
    \centering
    \begin{subfigure}[b]{0.3\textwidth}
        \includegraphics[width=\textwidth]{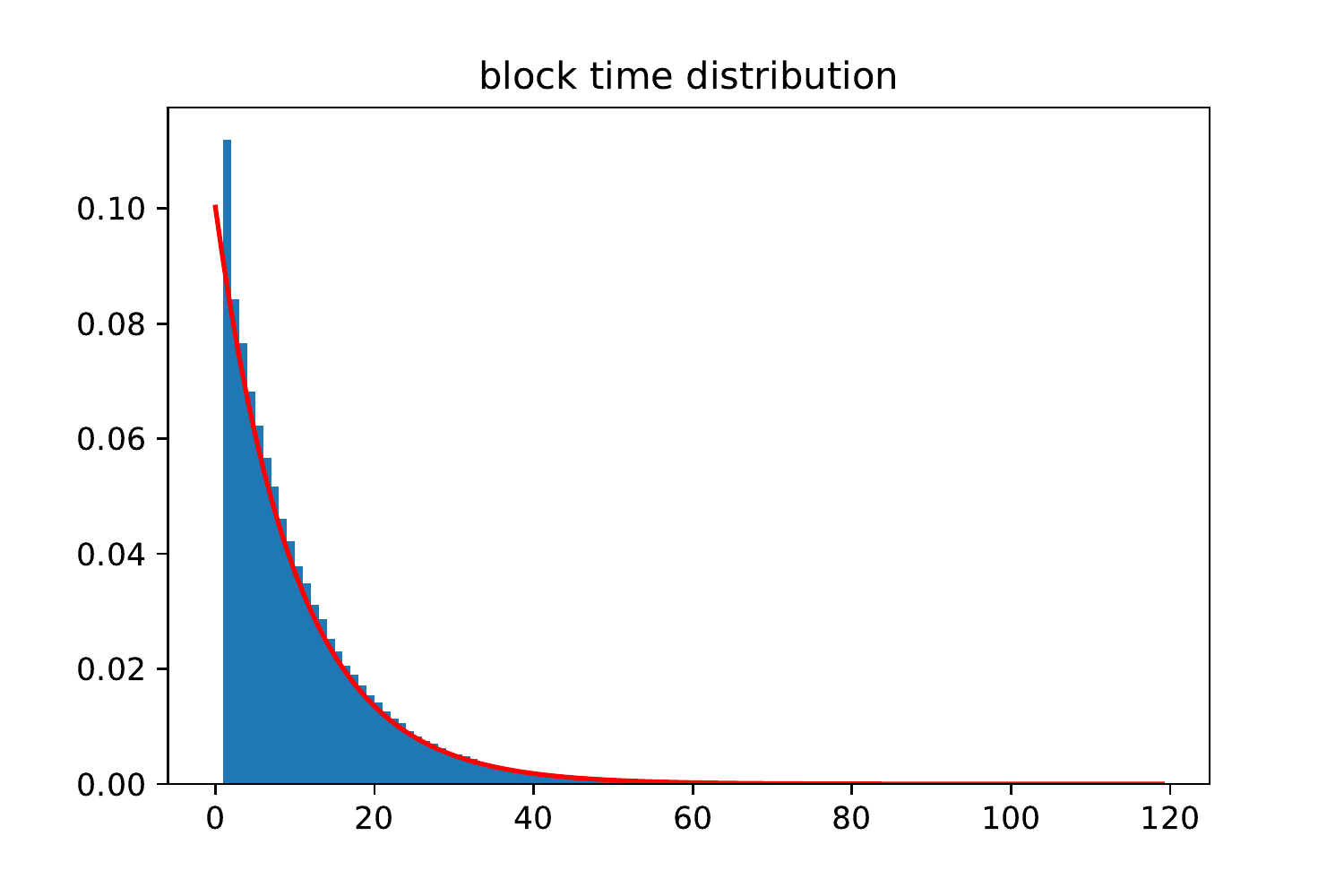}
        \caption{The time distribution between consecutive blocks.}
    \end{subfigure}
    \begin{subfigure}[b]{0.3\textwidth}
        \includegraphics[width=\textwidth]{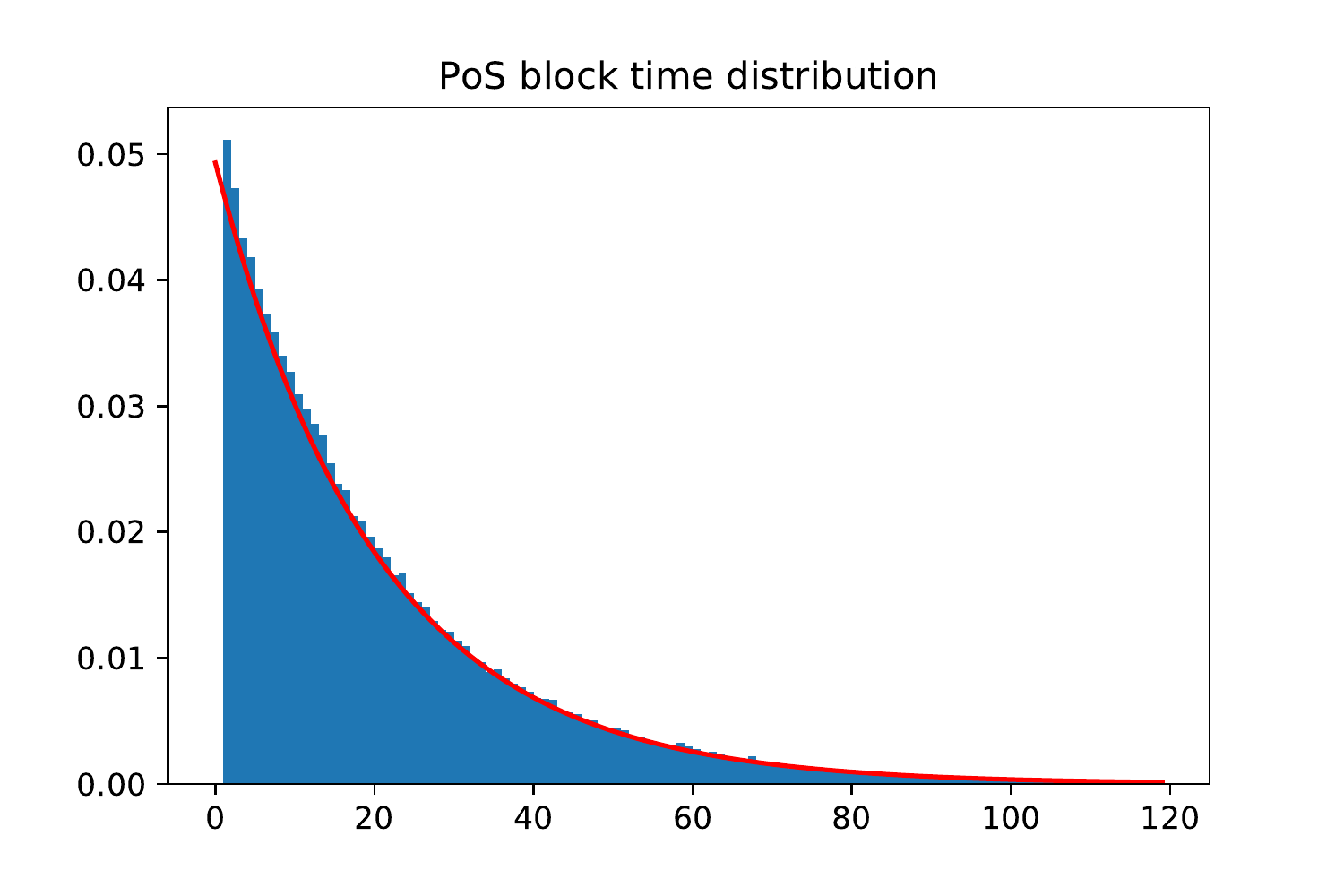}
        \caption{The time distribution between consecutive PoS blocks.}
    \end{subfigure}
    \begin{subfigure}[b]{0.3\textwidth}
        \includegraphics[width=\textwidth]{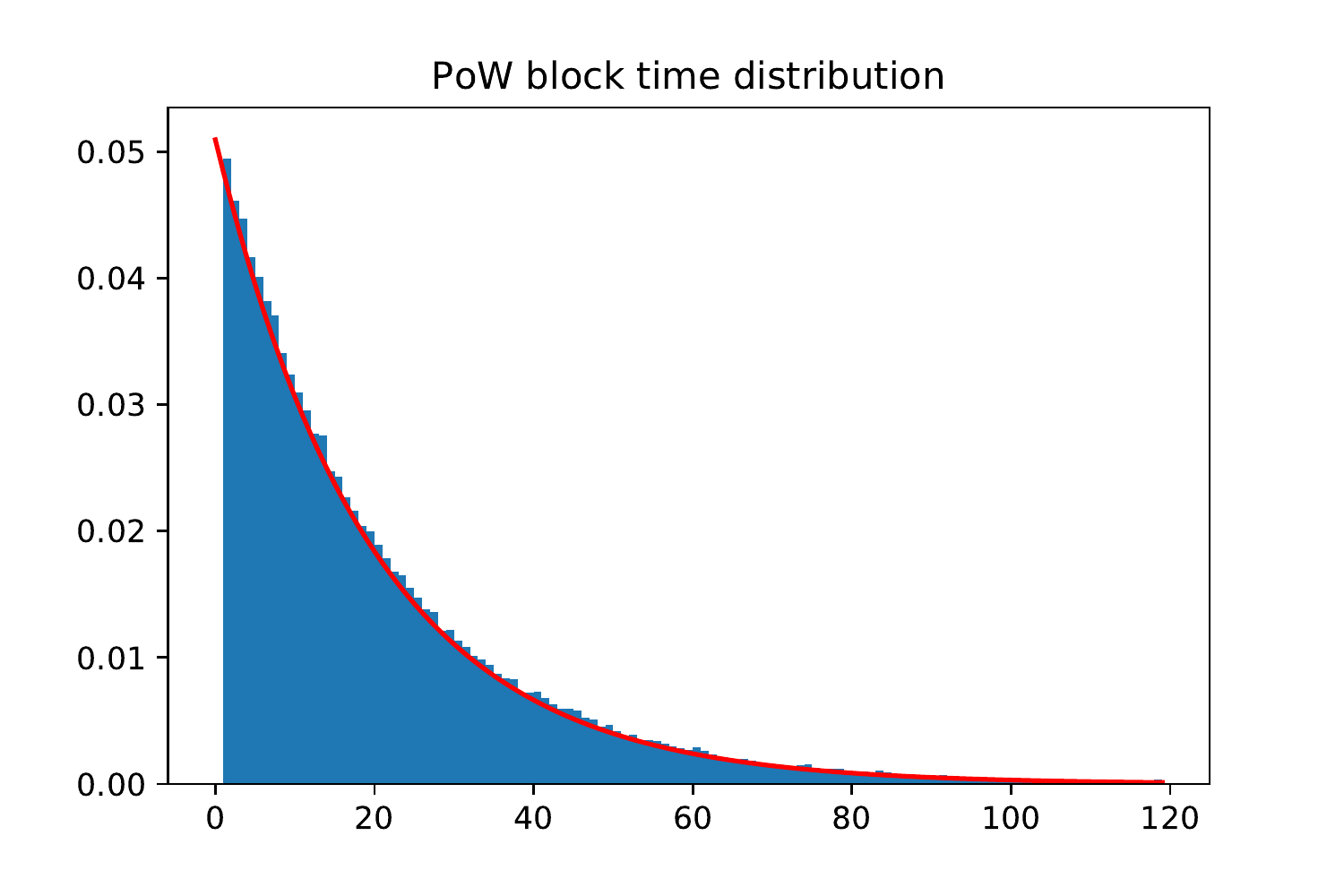}
        \caption{The time distribution between consecutive PoW blocks.}
    \end{subfigure}
    \caption{PoS/PoW block time distribution. The bars represent the block time density and the line is the exponential distribution fit.}
    \label{fig:block-time}
\end{figure}

The block time statistics are presented in \cref{table:block-time} and \cref{fig:block-time}. The distribution of PoS/PoW/All block time fits well the exponential distribution, as shown in \cref{fig:block-time}. The corresponding average rate are $\sum_{s \in S}{s} / d_s$, $\sum_{m \in M}{m} / d_m$, and $\sum_{s \in S}{s} / d_s + \sum_{m \in M}{m} / d_m$.

\begin{figure}[h]
    \centering
    \includegraphics[width=0.45\textwidth]{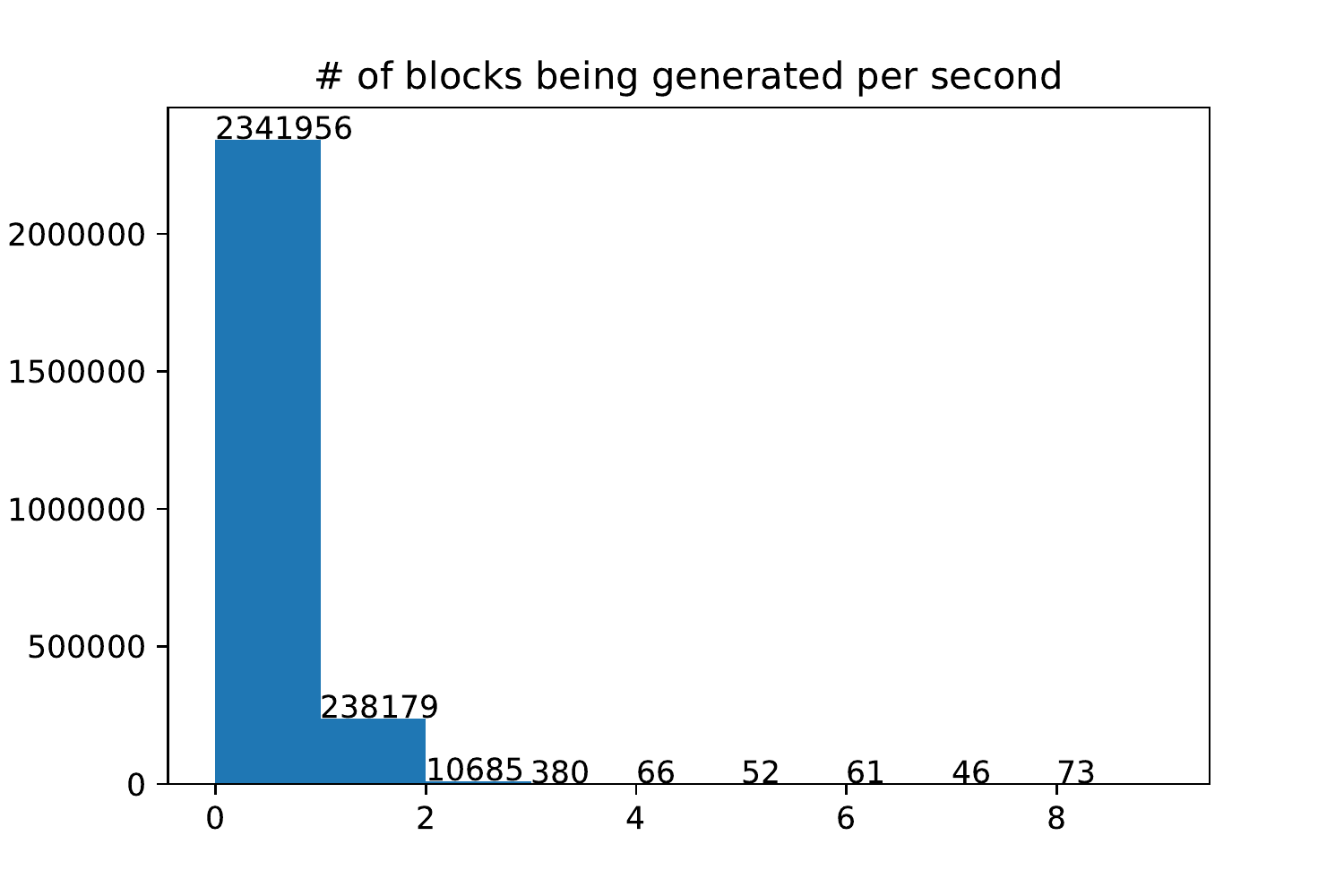}
    \caption{Number of blocks per second distribution.}
    \label{forks}
\end{figure}

Additionally, the number of blocks being generated at each second is demonstrated in \cref{forks}. The fraction of orphan blocks is approximately 4.4\%. However, this assumes a perfect connection between all miners and stakers; in reality, a higher rate is expected.

\begin{table}
\centering
\begin{tabular}{@{}rrr@{}}
\toprule
& Mean & Standard deviation \\ \midrule
All blocks & 10.366 & 10.050 \\
PoS blocks & 20.759 & 20.575 \\
PoW blocks & 20.705 & 20.420 \\
\bottomrule

\end{tabular}
\caption{The mean and standard deviation of block time.}
\label{table:block-time}
\end{table}

\subsection{Difficulty Adjustment}

\begin{figure}
    \centering
    \begin{subfigure}[b]{0.45\textwidth}
        \includegraphics[width=\textwidth]{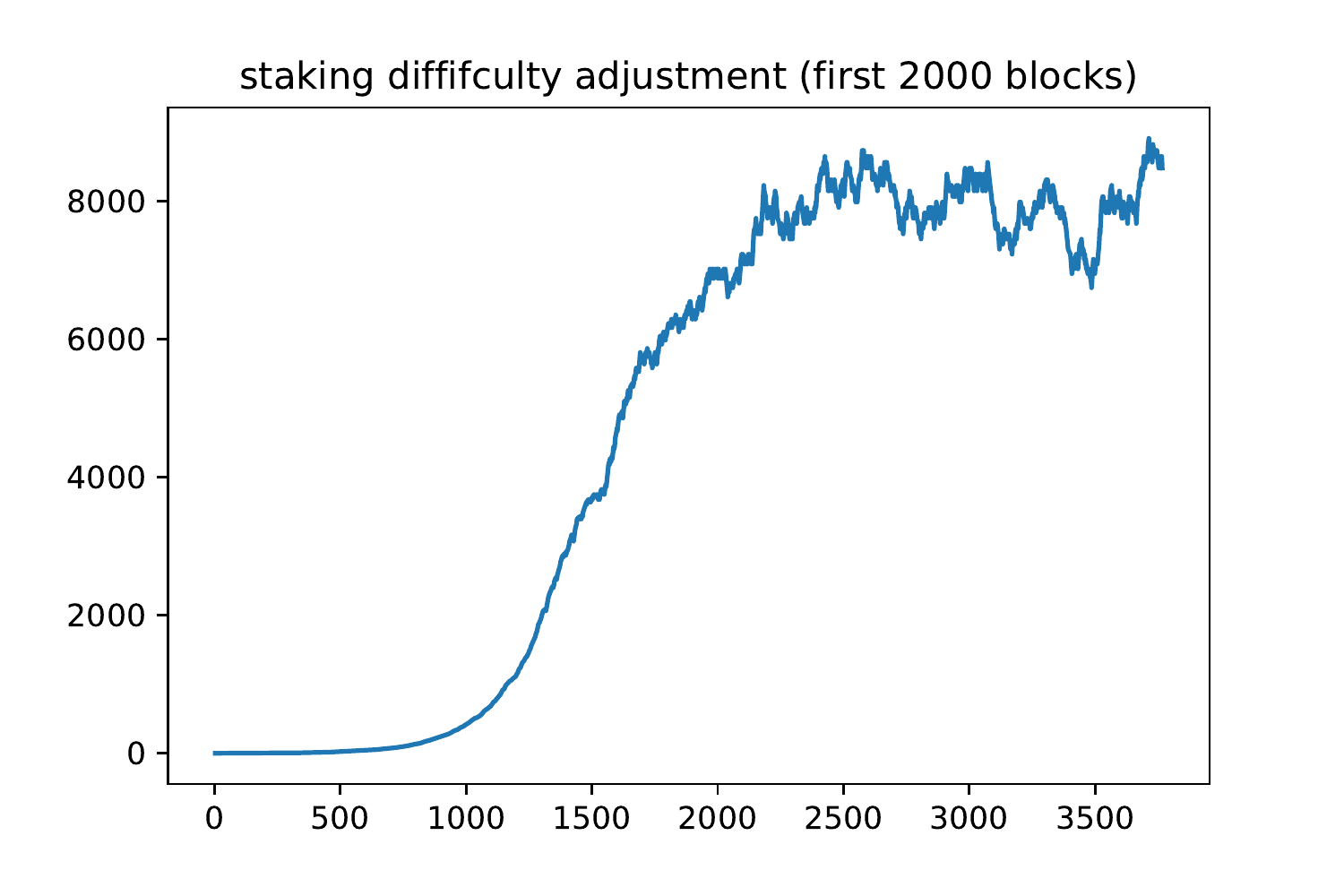}
        \caption{Staking difficulty $d_s$.}
    \end{subfigure}
    \begin{subfigure}[b]{0.45\textwidth}
        \includegraphics[width=\textwidth]{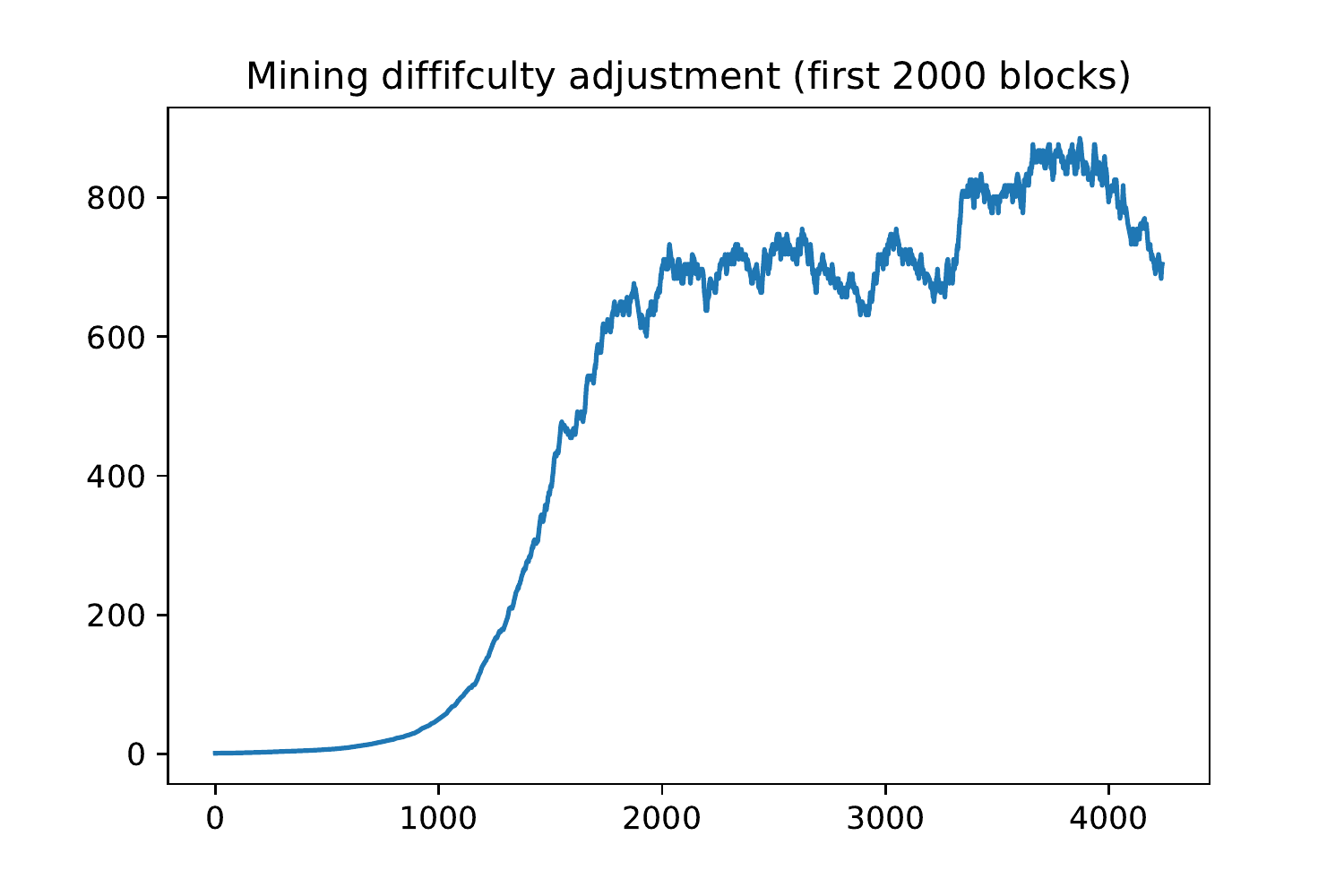}
        \caption{Mining difficulty $d_w$.}
    \end{subfigure}
    \caption{The staking/mining difficulties of the first 2000 PoS/PoW blocks.}
    \label{fig:diff-adj}
\end{figure}

For simulation purposes, we implemented a naive representation of the difficulty adjustment algorithm. The difficulty fluctuations are depicted in \cref{fig:diff-adj}. The algorithm gradually picks up the right mining/staking difficulty to match the current network hash/stake power, and then fluctuates within an acceptable range. The staking difficulty is approximately 10x the mining difficulty, which is expected as the total stake is 10x the hash rate.

\section{Conclusions And Future Work}

In this article, we have presented a novel consensus protocol, named Unity, for a public blockchain context. Unity combines two independent stochastic processes (mining and staking), into a coherent hybrid process. The simulation result indicates that the Unity protocol is fair to both miners and stakers. The time between consecutive PoW/PoS/All blocks follows the exponential distribution.

Different experiments and tests will be explored further in future publications and implementations. Future work concerns deeper analysis of the system stability with more complex miner and staker dynamics, a formal proof of the protocol over all the corner cases using game theory, and an economic model for Unity's implementation.

\clearpage
\begingroup
\raggedright
\bibliographystyle{alpha}
\bibliography{main}
\endgroup

\clearpage
\begin{appendices}
\section{Future Mining Attack}
\label{future_mining_attack}

Timestamp selection is closely related to the existence of a global clock, and therefore is very important in both types of block forging. The general rule being used today is that with some time bound $t_{future}$, nodes in the network following the protocol will refuse to accept or propagate blocks they see such that $t_{block} > \phi + t_{future}$.

Define $\alpha \in [0, 1] \subset R $ as the total forging power of the network, and define $\beta = 1-\alpha$ as the total forging power deviating from the standard strategy. Assume that the deviation of the strategy is of the form described in \cref{staking_algo} and \cref{mining_algo}, with the exception that when importing blocks from other sources, remove the check for $t_{block} \leq \phi + t_{future}$. Consequently, actors of this behaviour will mine blocks with timestamps arbitrarily far into the future. However, if we assume a simplification that difficulty algorithms adjust instantaneously to the network forging power. Assume $\alpha > 1/2$ it follows that $d_{main} > d_{deviate}$, therefore $td_{main} > td_{deviate}$ as more blocks are mined. The conclusion is that even if deviating allows you to generate blocks faster, the loss in total difficulty incurred ensures that by the time the main chain catches up to the future chain (recall that expected behaviour does not allow the acceptance of blocks until it is within a delta bound), it will have a higher total difficulty.

In the special case, assume that at time $t_0$ when the current highest block is $B_0$ and $B_0$ is known to every player we are going to introduce.

Assume there are PoS block producers Alice and Bob. Alice is eligible to produce a PoS block at $t_a$. Bob is eligible to produce a PoS block at $t_b$. How $t_a$ compares to $t_b$ is unknown to all other players. And we assume that $t_a$ is the earliest time among those who don't publish blocks early, and $t_b$ is the earliest time among those who publish future blocks early.

Assume there are PoW miners Charlie and David. Charlie and David are going to use different mining strategies. To compare their strategies fairly, assume that Charlie and David both have half the total hash power and they are very likely to produce a block at the same time ($\alpha = \beta$). Therefore the strategies of the 4 players look like the following:

\begin{itemize}
    \item Alice: produce and publish her PoS block $B_{s1a}$ at time $t_a$ with timestamp $t_a$
    \item Bob: produce and publish his PoS block $B_{s1b}$ at time $t_0$ with future timestamp $t_b$
    \item Charlie: always mine on the current highest weight block which is not from future (not with future timestamp)
    \item David: always mine on the highest weight block he sees (even with future time stamp)
\end{itemize}

Game begins from $t_0$, (\textbf{EVENT 0}):

\begin{itemize}
    \item \textbf{EVENT 0} $t_0$ : Bob produces and publishes his block $B_{s1b}$ with timestamp $t_b$. Charlie mines on $B_0$ while David mines on $B_{s1b}$.
    \item \textbf{EVENT 1} $t_x$: $t_x < t_a$ and $t_x < t_b$ (if $t_x$ is bigger than either $t_a$ or $t_b$, the game of future blocks doesn't exist anymore). Both Charlie and David produced a PoW block and a fork is created. Charlie's block $B_{w1c}$ is with timestamp $t_0$. David's block $B_{w1d}$ is with timestamp $t_b + 1$.
\end{itemize}

Now let's take a look at the weights of both Charlie's and David's chains:

\begin{itemize}
    \item Weight of Charlie's chain $W(chain_c) = W(B_0) + W(B_{w1c})$
    \item Weight of David's chain $W(chain_d) = W(B_0) + W(B_{s1b}) + W(B_{w1d})$
\end{itemize}

It is obvious to see that at $t_x$, $W(chain_d) > W(chain_c)$ despite that $chain_d$ includes blocks from future. Assume that Charlie does not switch to $chain_d$ because he is confident about his strategy.

\begin{itemize}
    \item \textbf{EVENT 2} $t_a$: Alice produces and publishes her block $B_{s1a}$ with timestamp $t_a$. $B_{s1a}$ references to the highest block $B_{w1c}$.
\end{itemize}

Note here that $B_{s1a}$ cannot reference to $B_{w1d}$ because $B_{s1a}$ conflicts with $B_{s1b}$. The PoS block $B_{s2b}$ from the next PoS round of Bob's block is not likely to come out before $t_a$ even $t_b < t_a$. (We have assumed that $t_a$ is the earliest time among those who don't publish future blocks, so $t_a$ is quite an early time). If Charlie and David both produced another (or more) PoW blocks before $t_a$, it doesn't change the weight comparison. After $t_a$, $chain_c$ and $chain_d$ are at the same conditions again: same block height, same hash power, same stake power (a new round of PoS starts nearly the same time at $t_a$ and $t_b$).

Now we take a look again at the weights of both $chain_c$ and $chain_d$:

\begin{itemize}
    \item $W(chain_c) = W(B_0) + W(B_{w1c}) + W(B_{s1a})$
    \item $W(chain_d) = W(B_0) + W(B_{s1b}) + W(B_{w1d})$
\end{itemize}

The conclusion of the above analysis is that: at time $t_a$, the weight of a chain depends on the timestamp (or the eligible time) of the PoS block but not on how early it is published. There is no proof that Bob and David's future block strategy is better than Alice and Charlie's. Besides, Alice gains more transactions fee by waiting until $t_a$ and including more transactions in her block; Charlie can be more confident about his block as his block has an earlier timestamp.

\section{Slashing Mechanisms}
\label{appendix:slashing}

\todo{Edit this section}

To discourage stakers from working on multiple branches when a fork comes into existence, a slashing mechanism can be introduced. To our best knowledge, there are two types of slashing mechanisms: slashing with hard evidence or slashing without hard evidence.

\subsection{Slashing with hard evidence}

Slashing with hard evidence is to punish only when a proof can be constructed to prove that a staker has deviated from the protocol on purpose. The rules are:
\begin{itemize}
    \item To punish a staker who has produced more than one block for the same block number;
    \item To punish a staker who has produced a block on top of a chain of weight $w_1$ with timestamp $t_1$, and another block on top of a chain of weight $w_2$ with timestamp $t_2$, where $w_1 >= w_2$ and $t_1 <= t_2$.
\end{itemize}

\subsection{Slashing without hard evidence}

Slashing without hard evidence to slash any staker for producing a block on a side chain, from the main chain, for example the Dunkle \cite{ethereum2}. It eliminates the undetectable nothing-at-stake problem, described in \cite{brown2018formal}.  Accounts are punished on chain $C$ for having produced blocks on other chains. The amount of punishment $S$ should be greater than block reward $R$ so that the expectation of revenues $E = R - S < 0$.

If we assume, under a certain network condition, the probability of accidentally producing an orphaned block is $a$ ($0 < a < 1$) and we have $S = n \cdot R$. When a staker is eligible to produce a block, his expected gain is $(1 - a) \cdot R$ while his expected lose is $a \cdot S$. For his expected gain to be greater than his expected lose, we have $(1 - a) \cdot R > a \cdot S  \geq   n < (1 - a) / a$. For instance, the current orphaned block rate of the Ethereum network is around 10.13\% (866869 uncle blocks out of 8555426 total blocks), then we have $n < 8.87$.

\subsection{Public Double-Spend Attack}

Public double-spend attack is different from the private double-spend attack, in a sense that the attacker constructs a parallel chain in public rather than in private. In the hybrid PoW/PoS context, an  attacker with over 50\% hash power builds their chain in public and wants the stakers to work on both the main chain and his side chain (if no punishment is enforced for stakers). The attacker's strategy is to always publish his block after a honest miners' block. Once his double-spending transaction is accepted, he always publish his blocks whenever mined; with high probability, his side chain will be the main chain eventually. 

Specifically, honest miners mine a block $B_{wh}$ while a malicious miner (attacker) mines a block $B_{wm}$ and publishes it after $B_{wh}$. Then Alice, the earliest eligible PoS block producer, builds a PoS block $B_{pa}$ on top of $B_{wh}$ because she saw it first. Then another PoS block producer Bob decides to build a block $B_{pb}$ on top of $B_{wm}$ to compete with Alice's main chain despite that Bob's block weights less than Alice's. The chain is now forked and the two forks grow in the same pace under the control of the attackers, assuming all stakers just support both forks by producing their PoS blocks whenever eligible. Several blocks later, when attackers are ready, they just produce and publish a PoW block earlier than honest miners, so that attackers' chain becomes the main chain (with higher total weight) and honest miners will switch to it. Then, the fork ends and attackers have successfully double spent.

The key solution is to convince stakers to not support the attacker's chain. After the side chain is maintained for a while, it becomes obvious to everyone in the network that, this side chain is very likely mined by a mining group with over 50\% hash power, because normally miners will not mine on a side chain that keeps up with the main chain for so long time unless they are attackers who are confident with their over 50\% hash power.

In this situation, a slashing mechanism like Dunkle does stop stakers from supporting both chains but it also forces stakers to make a choice between the main chain and the attacker's chain, because the stakers who supported the losing chain will eventually be punished. So the question is how many stakers will be convinced, either in public or private, by attackers to switch to attackers' chain:

\begin{itemize}
    \item If the majority of the stakers do care about the immutability of the network, they should refuse to support attackers' side chain.
    \item If the majority of the stakers only care about block rewards, they should probably support attackers' side chain if attackers showed enough hash power strength. With the support of the persuaded stakers, the attacker can achieve $hash\ power + stake\ power > 100\%$. Although this is beyond the designed protocol tolerance, such collusion should be made difficult to take place.
\end{itemize}

Based on the above analysis, applying slashing without hard evidence could be a double-edged sword:
\begin{itemize}
    \item Advantage: it stops stakers from supporting both forks and increases the difficulty and risky of collusion (it is harder for an attacker to prove their hash power in early stage of the fork without enough support of stakers).
    \item Side effect: if the attacker's chain wins (after successfully convincing stakers to collude and a group with $hash\ power + stake\ power > 100\%$ is formed), the stake of honest stakers will be slashed. This creates another incentive for miners and stakers to collude. Additionally, stakers may be encouraged to join a stake pool to reduce the risk of staking on a side chain.
\end{itemize}

\subsection{Alternatives to Slashing}

An alternative to slashing is to delay the finality of conflicting transactions (in one branch but not the other) at the application level. Unlike the private double-spend attack, the public attack pattern exposes those double spending transactions to the public. Exchanges, explorers and wallets could detect such pattern and delay the finality of such transactions.

\end{appendices}

\end{document}